\documentclass[twocolumn,amsmath,amssymb,aps,prd,nofootinbib]{revtex4}

\usepackage{graphicx}
\usepackage{color}

\begin{document}

\title{Unification of inflation, dark energy, and dark matter within the Salam-Sezgin cosmological model}

\author{Alfredo B. Henriques}

\email{alfredo@fisica.ist.utl.pt}

\affiliation{Centro Multidisciplinar de Astrof\'{\i}sica -- CENTRA and
Departamento de F\'{\i}sica, Instituto Superior T\'ecnico, UTL, Av.\ Rovisco
Pais, 1049-001 Lisboa, Portugal}
\author{Robertus Potting}

\email{rpotting@ualg.pt}

\affiliation{Centro Multidisciplinar de Astrof\'{\i}sica -- CENTRA and
Departamento de F\'{\i}sica, FCT, Universidade do Algarve, Campus de Gambelas,
8005-139 Faro, Portugal}

\author{Paulo M. S\'a}

\email{pmsa@ualg.pt}

\affiliation{Departamento de F\'{\i}sica, FCT, Universidade do
Algarve, Campus de Gambelas, 8005-139 Faro, Portugal}

\date{March 11, 2009}

\begin{abstract}
We investigate a cosmological model, based on the Salam-Sezgin
six-dimensional supergravity theory and on previous work by
Anchordoqui, Goldberg, Nawata, and Nu\~nez. Assuming a period of
warm inflation, we show that it is possible to extend the
evolution of the model back in time, to include the inflationary
period, thus unifying inflation, dark matter, and dark energy
within a single framework. Like the previous authors, we were not
able to obtain the full dark matter content of the Universe from
the Salam-Sezgin scalar fields. However, even if only partially
successful, this work shows that present-day theories, based on
superstrings and supergravity, may eventually lead to a
comprehensive modelling of the evolution of the Universe. We find
that the gravitational-wave spectrum of the model has a
non-constant negative slope in the frequency range
$(10^{-15}-10^{6})\,\mbox{rad/s}$, and that, unlike standard
(cold) inflation models, it shows no structure in the MHz/GHz
range of frequencies.
\end{abstract}


\maketitle

\section{Introduction}

We still are a long way from having a theoretically sound unified
model explaining both the inflationary epoch and the standard era
of expansion. The problem has become even more complex since the
discovery of the dark energy content and the resulting present
accelerated expansion of the
Universe~\cite{riess-et-al,perlmutter-et-al}. Recently,
Anchordoqui, Goldberg, Nawata, and
Nu\~{n}ez~\cite{anchordoqui-et-al} proposed a very interesting
model based on the Salam-Sezgin six-dimensional supergravity
model~\cite{salam-sezgin}, whose spontaneous compactification from
six to four dimensions gives rise to a potential, henceforth
called the Salam-Sezgin potential. When this model is lifted to
M-theory, the internal space is found to be
non-compact~\cite{cvetic-et-al}, circumventing existing no-go
theorems and allowing the potential to be positive. The work of
Anchordoqui \emph{et al.}~\cite{anchordoqui-et-al} is limited to
the investigation of the epochs subsequent to primordial
nucleosynthesis. The authors derived a solution of the Einstein
field equations in qualitative agreement with the observations of
an accelerating universe, where the dark energy comes from a
scalar field slowly rolling down the (exponential) Salam-Sezgin
potential. Another important result was the finding that the model
also contains a cold dark matter component ($p=0)$, coming from
another scalar field interacting with the dark energy field, whose
particles, as a result of this interaction, have an effective mass
slowly evolving with the dark-energy scalar field. Unfortunately,
dark matter from the Salam-Sezgin scalar field only accounts for
up to about 6\% of the total matter content of the universe.

In the present paper we extend this model back in time. It has
been known for a long time~\cite{lucchin-matarrese} that an
exponential potential of the generic form  $V=V_{0}\exp
(-\sqrt{8\pi G} \gamma \phi)$ is able to support inflation only if
$\gamma < \sqrt{2}$. The Salam-Sezgin potential fails to obey this
requirement, having $\gamma = \sqrt{2}$. But, if we assume a
period of warm inflation~\cite{berera}, it is possible to show
that the Salam-Sezgin potential is able to drive enough inflation.
We obtain, therefore, within the Salam-Sezgin cosmological model,
a triple unification of inflation, dark matter, and dark energy.
As in Ref.~\cite{anchordoqui-et-al}, we were not able to obtain
the full dark matter content of the Universe from the Salam-Sezgin
scalar fields, the rest of the dark matter and the baryonic matter
having to be put in by hand. Even if only partially successful, we
believe that our analysis shows that present day theories, based
on supergravity and superstrings, may eventually be successful in
providing a comprehensive description of the evolution of the
Universe.

During the period of warm inflation, energy is continuously
transferred from the Salam-Sezgin scalar fields to a radiation
fluid due to the presence of dissipative terms in the energy
conservation equations~\cite{berera} (for a review see
\cite{warm-inflation} and references therein). For illustrative
purposes we will assume a simple phenomenological dissipative
term, following the suggestion of Yokoyama and
Maeda~\cite{yokoyama-maeda}. In this work we will not attempt to
justify the occurrence of warm inflation, which should be due to
the coupling of the scalar fields to a heat bath of background
fields~\cite{berera-kephart}. We just mention that plenty of such
fields are present if the model is assumed to be derived from
string theory.

In the next section we briefly review the Salam-Sezgin model. In
Sect.~\ref{sect-the-evolution-model} we derive the equations
defining the model and introduce the details of how, following the
inflationary period, we calculate the amount of dark energy and
dark matter. This is followed by a section on the numerical
simulations of the evolution of the universe, since the beginning
of the inflationary period till the present time, and the analysis
of the results obtained. Section~\ref{sect-gravitational-waves} is
dedicated to the calculation of the full energy spectrum of the
gravitational waves generated during the expansion of the
universe, as it would be measured today by an ideal detector.
Finally, we end with the summary and the conclusions.

\section{The Salam-Sezgin potential\label{sect-salam-sezgin_potential}}

We begin with a review of the Salam-Sezgin model, following the notation of
Anchordoqui \emph{et al.}~\cite{anchordoqui-et-al}. The Salam-Sezgin model
represents a six-dimensional supergravity~\cite{salam-sezgin}. It has been
shown by Cvetic, Gibbons, and Pope~\cite{cvetic-et-al} that this model can be
uplifted consistently to type~I superstring theory. Its bosonic part is given
by the action
\begin{eqnarray}
S &=& \frac{1}{4\kappa^2} \int d^6x \sqrt{g_6} \bigg[ R - \kappa^2
( \partial_{M} \sigma )^2 -\frac{2g^2}{\kappa^2} e^{-\kappa\sigma}
 \nonumber
\\
&&  - \kappa^2 e^{\kappa\sigma} F_{M\!N}^2 - \frac{\kappa^2}{3}
e^{2\kappa\sigma} G_{M\!N\!P}^2 \bigg]. \label{ss}
\end{eqnarray}
Here, $g_6=\det g_{M\!N}$, $R$ is the Ricci scalar of $g_{M\!N}$,
$\sigma$ is a scalar field, $F_{M\!N}=\partial_{[M} A_{N]}$,
$G_{M\!N\!P}=\partial_{[M}B_{N\!P]} + \kappa A_{[M} F_{N\!P]}$,
where $A_N$ is a gauge field and $B_{N\!P}$ is the Kalb-Ramond
field; capital Latin indices run from $0$ to $5$. Introducing $G_6
\equiv 2 \kappa^2$ and $\xi \equiv 4g^2$ and rescaling $\phi
\equiv -\kappa \sigma$ leads to
\begin{eqnarray}
S &=& \frac{1}{2 G_6} \int d^6x \sqrt{g_6} \bigg[ R -
(\partial_M \phi)^2 - \frac{\xi}{G_6} e^\phi \nonumber \\
&& - \frac{G_6}{2} e^{-\phi} F_{M\!N}^2 - \frac{G_6}{6} e^{-2\phi}
G_{M\!N\!P}^2 \bigg] . \label{action}
\end{eqnarray}
Here the length dimensions are: $[G_6]=L^4$, $[\xi]=L^2$,
$[\phi]=[g_{M\!N}^2]=1$, $[A_M^2]=L^{-4}$, and
$[F_{M\!N}^2]=[G_{M\!N\!P}^2]= L^{-6}$.

Salam and Sezgin~\cite{salam-sezgin} showed that Lagrangian~(\ref{ss}) can be
compactified on the direct product $M = M_4 \times S_2$, where
$M_4\equiv(\mbox{Minkowski})_4$.  It was shown by Gibbons, G\"uven, and
Pope~\cite{gibbons-et-al} that this is the unique ground state among all
non-singular solutions with a four-dimensional Poincar\'e, de Sitter or
anti-de Sitter symmetry. The monopole configuration they found has the metric
on $M$ locally of the form
\begin{eqnarray}
ds_6^2 = ds_4(t,{\vec x})^2 + e^{2f(t,{\vec x})} r_c^2
(d\vartheta^2 +\sin^2\vartheta d\varphi^2), \label{metric}
\end{eqnarray}
where $(t, \vec x)$ denotes a local coordinate system on $M_4$ and
$r_c$ is the compactification radius of $S_2$. The scalar field
$\phi$ is taken to depend only on the point of $M_4$, i.e.,
$\phi=\phi(t, \vec x)$. The gauge field $A_M$ is excited on $S_2$
and is of the form
\begin{eqnarray}
A_\vartheta=0,\qquad A_\varphi=b\cos \vartheta, \label{FC}
\end{eqnarray}
which satisfies the Maxwell equations (obtained by varying $A_M$)
and yields the field strength
\begin{eqnarray}
F_{M\!N}^2 = 2b^2 e^{-4f}/r_c^4. \label{FS}
\end{eqnarray}
Finally $B_{N\!P}$ is taken to vanish. It follows that also
$G_{M\!N\!P} = 0$.

The Ricci scalar can now be written as~\cite{wald}
\begin{eqnarray}
R[M] = R[M_4] + e^{-2f} R[S_2] - 4\Box f - 6(\partial_\mu f)^2,
\end{eqnarray}
where $R[M]$, $R[M_4]$, and $R[S_2]$ denote the Ricci scalars of
the manifolds $M$, $M_4$, and $S_2$, respectively. It follows
\begin{eqnarray}
R[S_2] = 2/r_c^2 \label{007}
\end{eqnarray}
and $\sqrt{g_6} = e^{2f}r_c^4 \sqrt{g_4}$, where $g_4 = \det g_{\mu\nu}$ (with
Greek indices running from 0 to 3). We now define the gravitational constant
in four dimensions as
\begin{eqnarray}
G_4\equiv \frac{1}{m_{\textsc{p}}^2} = \frac{G_6}{32 \pi^2 r_c^2}.
\end{eqnarray}
It follows that by using the Salam-Sezgin monopole configuration
we can re-write the action in Eq.~(\ref{action}) as follows
\begin{eqnarray}
S &=& \frac{1}{ 16\pi G_4 } \int d^4 x \sqrt{g_4} \Bigg\{ e^{2f}
\bigg[
R[M_4] + \frac{2}{r_c^2}e^{-2f} + 2(\partial_\mu f)^2 \nonumber \\
&&  - (\partial_\mu \phi)^2 \bigg]  - \frac{\xi}{G_6} e^{2f+\phi} - \frac{G_6
b^2}{r_c^4}\, e^{ -2f-\phi} \Bigg\}. \label{action4string}
\end{eqnarray}
Using the rescaling of the metric $\hat g_{\mu\nu}\equiv e^{2f}
g_{\mu\nu}$ and $\sqrt{\hat{g}_4}=e^{4f} \sqrt{g_4}$, the model is
taken into the Einstein conformal frame where the action
Eq.~(\ref{action4string}) takes the form
\begin{eqnarray}
S &=& \frac{1}{ 16\pi G_4  } \int d^4 x \sqrt{\hat{g}_4} \bigg[
R[\hat{g}_4] - 4 (\partial_\mu f)^2 - (\partial_\mu \phi)^2
\nonumber \\
&& - \frac{\xi}{G_6} e^{-2f+\phi} - \frac{G_6 b^2}{r_c^4}
e^{-6f-\phi} + \frac{2}{r_c^2} e^{-4f}\bigg]. \label{action4}
\end{eqnarray}
The four-dimensional Lagrangian is then
\begin{eqnarray}
L = \frac{\sqrt{g}}{ 16\pi G  } \left[ R - 4 (\partial_\mu f)^2 -
(\partial_\mu \phi)^2 - V(f,\phi)  \right], \label{finallagran}
\end{eqnarray}
with
\begin{eqnarray}
V(f,\phi) \equiv \frac{\xi}{G_6} e^{-2f+\phi} + \frac{G_6
b^2}{r_c^4} e^{-6f-\phi} - \frac{2}{r_c^2} e^{-4f},
\end{eqnarray}
where we have written $g\equiv\hat{g}_4$, $R\equiv R[\hat{g}_4]$
and $G\equiv G_4$.

Following Anchordoqui \emph{et al.}~\cite{anchordoqui-et-al}, we
now perform the field redefinition
\begin{eqnarray}
x \equiv (\phi+2f)/\sqrt{16\pi G}, \qquad y \equiv
(\phi-2f)/\sqrt{16\pi G}.
\end{eqnarray}
The Lagrangian becomes
\begin{eqnarray}
L = \sqrt{g} \left[ \frac{R}{16\pi G} - \frac{1}{2} (\partial x)^2
- \frac{1}{2}(\partial y)^2 - V(x,y)  \right], \label{ll}
\end{eqnarray}
with the Salam-Sezgin potential given by
\begin{eqnarray}
V(x,y)=A_1 e^{\alpha y} \left( 1-2A_2 e^{-\alpha x} +A_3
e^{-2\alpha x} \right), \label{potencial}
\end{eqnarray}
where we have introduced the notation
\begin{eqnarray}
A_1=\frac{\xi m_{\textsc{p}}^4}{512 \pi^3 r_c^2}, \quad
A_2=\frac{32\pi^2}{\xi m_{\textsc{p}}^2}, \quad
A_3=\frac{1024\pi^4 b^2}{\xi m_{\textsc{p}}^4},
\end{eqnarray}
and $\alpha=\sqrt{16\pi G}=4\sqrt{\pi}/m_{\textsc{p}}$.

\section{The evolution model\label{sect-the-evolution-model}}

In our analysis of the Salam-Sezgin cosmological model, we will
divide the evolution of the universe in two stages. The first
stage starts at the beginning of the inflationary period and
extends well into the radiation-dominated epoch. During this first
stage of evolution, energy is continuously transferred from the
scalar fields $x$ and $y$ to a radiation fluid due to the presence
of dissipative terms in the energy conservation equations.
Inflation, which we assume to be of the warm type~\cite{berera},
is driven by the field $x$. When the energy density of radiation
is of the order $(10^{14}\,\mbox{GeV})^4$ (see
Sect.~\ref{sect-numerical-simulations} for details), we assume
that the energy transfer from the fields $x$ and $y$ to radiation
ceases. This marks the end of the first stage of evolution. During
the second stage, which extends up to the present epoch, the
$x$-field oscillates around the minimum of the potential, behaving
as cold dark matter with varying mass, as explained below. The
$y$-field, which behaves like dark energy, is practically constant
during most of the second stage of evolution and becomes dominant
in a recent epoch, in agreement with observations. In what follows
we present the relevant equations for both stages of evolution and
show how an unification of inflation, dark matter, and dark energy
can be achieved within the Salam-Sezgin cosmological model.

For a flat universe, described by the Friedmann-Robertson-Walker metric,
\begin{eqnarray}
 ds^2= -dt^2+a^2(t) d\chi^2,
\end{eqnarray}
where $a(t)$ is the scale factor and $d\chi^2$ is the metric of
the three-dimensional Euclidean space, the Einstein equations are
given by
\begin{eqnarray}
 \frac{\ddot{a}}{a} &=& -\frac{\alpha^2}{6} \left[ \dot{x}^2 +
\dot{y}^2 - V(x,y) + \rho_r \right],\label{adotdot1}
\\
 \left( \frac{\dot{a}}{a} \right)^2 &=& \frac{\alpha^2}{6} \left[
\frac{\dot{x}^2}{2} + \frac{\dot{y}^2}{2} + V(x,y) + \rho_r
\right], \label{friedmann1}
\end{eqnarray}
where $V(x,y)$ is the Salam-Sezgin potential given by
Eq.~(\ref{potencial}), $\rho_r$ is the energy density of a
radiation fluid with equation of state $p_r=\rho_r/3$, and a dot
denotes a derivative with respect to the comoving time $t$.

Let us assume that, during the first stage of evolution, energy is
continuously transferred from the scalar fields $x$ and $y$ to the
radiation fluid $\rho_r$ due to the presence of dissipative terms.
In this case, the energy conservation equations for the scalar
fields and the radiation fluid are
\begin{eqnarray}
 \ddot{x} + 3 \frac{\dot{a}}{a} \dot{x} +\frac{\partial V}{\partial
x} &=& - \Gamma_{x} \dot{x}, \label{x1a}
\\
 \ddot{y} + 3 \frac{\dot{a}}{a} \dot{y} +\frac{\partial
V}{\partial y} &=& - \Gamma_{y} \dot{y}, \label{y1a}
\\
 \dot{\rho}_r + 4\frac{\dot{a}}{a} \rho_r &=& \Gamma_x \dot{x}^2+
\Gamma_y \dot{y}^2, \label{rho1}
\end{eqnarray}
where $\Gamma_x$ and $\Gamma_y$ are the dissipative coefficients.

The form of the dissipative coefficients has been discussed in
literature and can be, in realistic models, quite complicated. For
illustrative purposes we will consider in this article simple
phenomenological dissipative coefficients, first introduced by
Yokoyama and Maeda~\cite{yokoyama-maeda}, of the form
\begin{eqnarray}
&& \Gamma_x(x,y)=f_x\sqrt{\frac{\partial^2V}{\partial
x^2}}, \\
&& \Gamma_y(x,y)=f_y\sqrt{\frac{\partial^2V}{\partial y^2}},
\end{eqnarray}
where $f_x$ and $f_y$ are free parameters. For an appropriate choice of these
parameters, a significant amount of energy is transferred from the scalar
fields $x$ and $y$ to the radiation fluid and, as a consequence, the energy
density of radiation is not diluted during inflation, as in the standard
(cold) inflationary models, and its influence in the evolution of the universe
is such that enough inflation occurs, despite the steepness of the
Salam-Sezgin potential.

To be general, let us assume that, immediately prior to the
inflationary phase, the universe may have, in addition to the
energy contained in the $x$ and $y$ scalar fields, an important
contribution from the radiation fluid. Eventually, the potential
energy $V(x,y)$ becomes dominant and warm inflation begins.
Initially, the energy density of radiation, $\rho_r$, decreases
faster than the energy density of the fields $x$ and $y$,
$\rho_{xy}=\dot{x}^2/2+\dot{y}^2/2+V(x,y)$. However, because of
the energy transfer due to the dissipative terms, $\rho_r$ quickly
reaches a state during which it decreases slower than $\rho_{xy}$.
At a certain point, $\rho_r$ becomes greater than $\rho_{xy}$ and
a smooth transition from warm inflation to a radiation-dominated
universe takes place~\cite{taylor-berera}.

During inflation the $x$-field slowly rolls down the potential
$V(x,y)$, approaching its minimum at $x_{\mbox{\scriptsize
min}}=(1/\alpha) \ln(A_3/A_2)$. During the inflationary period the
$y$-field does not play a significant role in the evolution of the
universe; it simply slowly rolls down the exponential (in the $y$
direction) potential.

Note that the potential (\ref{potencial}) can be expanded near its minimum,
yielding
\begin{eqnarray}
\hspace{-2mm} V(x,y)= A e^{\alpha y} +\frac12 M_x^2(x-x_{\mbox{\scriptsize
min}})^2+ ..., \label{potencial-expandido}
\end{eqnarray}
where
\begin{eqnarray}
M_x=\sqrt{2}\alpha\frac{A_1^{1/2}A_2}{A_3^{1/2}}e^{\alpha y/2}
\end{eqnarray}
is the time-dependent mass of the scalar field $x$ and
\begin{eqnarray}
A \equiv A_1 \left( 1-\frac{A_2^2}{A_3} \right)=
\frac{m_{\textsc{p}}^4}{512\pi^3 r_c b^2} \left( b^2\xi-1 \right). \label{A}
\end{eqnarray}
The first term on the right-hand-side of
Eq.~(\ref{potencial-expandido}) behaves like dark energy and, at
the present epoch, dominates the dynamics of evolution of the
universe. In order to achieve agreement with observations, the
constant $A$ should be very small, of the order of
$10^{-122}m_{\textsc{p}}^4$ (see
Sect.~\ref{sect-numerical-simulations} for details). This
extremely small, but different from zero, value means that
supersymmetry should to be broken within the Salam-Sezgin
cosmological model~\cite{anchordoqui-et-al}.

After the end of the inflationary period, energy transfer from the
$x$ and $y$ fields to the radiation fluid continues. As the
universe expands, its temperature drops and eventually the
interaction between the scalar fields and the thermal bath becomes
negligible. At this point, we turn the dissipative coefficients
$\Gamma_x$ and $\Gamma_y$ to zero. This marks the end of the first
stage of evolution. It is necessary to emphasize here, that the
residual amount of energy in the $x$-field at the end of the first
stage of evolution should be low, in order to make sure that the
universe undergoes a long enough radiation-dominated period,
ending well after
nucleosynthesis~\cite{liddle-urena,liddle-pahud-urena}. Our
numerical simulations (see next section) show that the transition
between the first and second stage of evolution occurs at
temperatures of the order of $10^{14}\,\mbox{GeV}$ and that the
scalar field $x$ is, at that time, of the order of $-10^{-17}
m_{\textsc{p}}$.

Due to the absence of the dissipative terms, the scalar field $x$
oscillates rapidly around its minimum during the second stage of
evolution. Since in the expanded potential
(\ref{potencial-expandido}) the dominant term is quadratic, the
oscillating $x$-field behaves like matter with equation of state
$p=0$ \cite{turner}. This pressureless matter behaves like cold
dark matter. Its energy density is proportional to the
$y$-dependent mass of the $x$-field (which is practically constant
during most of the second stage of evolution) and decreases as
$a^{-3}$,
\begin{eqnarray}
 c^2 \rho_x=C e^{\alpha y/2} \left( \frac{a_0}{a} \right)^3, \label{qaz}
\end{eqnarray}
where $c$ is the speed of light, $\alpha=4\sqrt{\pi G}/c$, and $a_0\equiv
a(t_0)$ is the value of the scale factor today\footnote{In the first stage of
evolution we use the natural system of units, with
$m_{\textsc{p}}=1/\sqrt{G}=1.22\times10^{19}\,\mbox{GeV}$, while in the second
stage we use the International Systems of Units.}. The constant $C$ should be
chosen in order to guarantee continuity of the solution at the transition from
the first to the second stage of evolution, namely,
\begin{eqnarray}
C= c^2 \rho_{x,\textsc{t}}  e^{-\alpha y_\textsc{t}/2} \left(
\frac{\rho_{r,0}}{\rho_{r,\textsc{t}}} \right)^{3/4}, \label{fix-C}
\end{eqnarray}
where the subscript $\textsc{t}$ refers to quantities evaluated at
the end of the first stage of evolution and we have taken into
account that
\begin{eqnarray}
\rho_r &=& \rho_{r,0} \left( \frac{a_0}{a} \right)^4,
\end{eqnarray}
where $\rho_{r,0} \equiv \rho_r(t_0)=4.6\times10^{-31}\mbox{
kg/m}^3$ is the density of radiation observed today.

Within the Salam-Sezgin cosmological model, the $x$-field accounts
for just a fraction of the total matter content of the
universe~\cite{anchordoqui-et-al}. Therefore, in order to be in
agreement with observational data, we introduce an additional
pressureless matter component, which accounts for other types of
dark matter and for the usual (baryonic) matter. For this
component, the energy conservation equation yields
\begin{eqnarray}
\rho_m &=& \frac{B}{c^2} \left( \frac{a_0}{a} \right)^3,
\end{eqnarray}
where $B$ is a constant.

The evolution equations for the $y$-field and the scale factor $a$
are now
\begin{eqnarray}
\ddot{y} &=& - 3\frac{\dot{a}}{a} \dot{y} - \alpha A e^{\alpha y}
-\frac{\alpha}{2} C  e^{\alpha y/2} \left( \frac{a_0}{a}
\right)^3, \label{ydotdot}
 \\
 \frac{\ddot{a}}{a} &=&
-\frac{\alpha^2}{12} \left[ 2\dot{y}^2 -2 A e^{\alpha y} + \left(
B + C  e^{\alpha y/2} \right) \left( \frac{a_0}{a} \right)^3
\right. \nonumber
 \\&& + \left. 2\rho_{r,0} c^2 \left( \frac{a_0}{a} \right)^4 \right],
\label{eq-adotdota}
 \\
 \left( \frac{\dot{a}}{a} \right)^2 \! \! &=& \frac{\alpha^2}{6} \left[
\frac{\dot{y}^2}{2} + A e^{\alpha y} + \left( B + C  e^{\alpha
y/2} \right) \left( \frac{a_0}{a} \right)^3 \right. \nonumber
 \\ && + \left. \rho_{r,0} c^2 \left( \frac{a_0}{a} \right)^4 \right].
\label{eq-adota}
\end{eqnarray}

In the above equations, the constants $A$ and $B$ should be chosen
consistently with observations, namely,
\begin{eqnarray}
&& Ae^{\alpha y_0} + \frac{\dot{y_0}^2}{2}= \rho_{\textsc{de},0} c^2,
\label{constraint-A}
\\
&& B+C e^{\alpha y_0/2}=\rho_{\textsc{m},0} c^2, \label{constraint-B}
\end{eqnarray}
where $\rho_{\textsc{de},0} = 6.9\times10^{-27} \mbox{ kg/m}^3$ and
$\rho_{\textsc{m},0}=2.6\times10^{-27} \mbox{ kg/m}^3$. To the values of
$\rho_{\textsc{de},0}$, $\rho_{\textsc{m},0}$, and $\rho_{r,0}$ used in this
article, correspond a value of the Hubble constant $H_0=71 \mbox{
km}\,\mbox{s}^{-1}\mbox{Mpc}^{-1}$.

Following Ref.~\cite{anchordoqui-et-al}, during the second stage
of evolution, we use, instead of the comoving time $t$, a new
variable $u$, defined as
\begin{eqnarray}
u=-\ln (1+z),
\end{eqnarray}
where $z=a_0/a-1$ is the redshift.

Using this new variable, the energy conservation equation for the
scalar field $y$ can be rewritten as
\begin{eqnarray}
y_{uu} &=& - \Bigg\{ \left[ \frac{\ddot{a}}{a} + 2 \left(
\frac{\dot{a}}{a} \right)^2 \right]  y_u + \alpha A e^{\alpha y}
\nonumber \\
&& +\, \frac{\alpha}{2}C e^{\alpha y/2}e^{-3u} \Bigg\}
\left(\frac{\dot{a}}{a}\right)^{-2}, \label{final-y}
\end{eqnarray}
where the subscript $u$ denotes a derivative with respect to $u$;
$\ddot{a}/a$ and $(\dot{a}/a)^2$ are functions of $u$, $y$ and
$y_{u}$ given by
\begin{eqnarray}
\hspace{-4mm}\frac{\ddot{a}}{a} &=& \frac{\alpha^2}{12} \left\{
4\alpha^2 \Big[ Ae^{\alpha y} + \left( B + C  e^{\alpha y/2}
\right)e^{-3u} \right.
\nonumber \\
&& + \left. \rho_{r,0} c^2
e^{-4u} \Big] y_u^2 (\alpha^2 y_u^2-12)^{-1} \right. + 2Ae^{\alpha y} \nonumber \\
\hspace{-4mm} && \left. -\left( B + C e^{\alpha y/2}
\right)e^{-3u} -2 \rho_{r,0} c^2 e^{-4u} \right\},
\label{alinhalinhaa}
\\
\hspace{-4mm} \left( \frac{\dot{a}}{a} \right)^2 &=& 2\alpha^2
\Big[ A e^{\alpha y}
+ \left( B + C  e^{\alpha y/2} \right)e^{-3u} \nonumber \\
\hspace{-4mm} && + \rho_{r,0} c^2 e^{-4u} \Big]
(12-\alpha^2y_u^2)^{-1}.
 \label{alinhaa2}
\end{eqnarray}

The set of equations Eqs.~(\ref{adotdot1})--(\ref{rho1}), for the first stage
of evolution, and (\ref{final-y})--(\ref{alinhaa2}), for the second stage,
will be solved numerically in the next section. The parameters $f_x$ and $f_y$
will be chosen such that enough inflation (65 or more $e$-folds) is achieved
during the first stage of evolution. The time of transition between the two
stages of evolution will be chosen such that the radiation-dominated epoch
taking place after inflation ends well after nucleosynthesis. The constant $C$
will be chosen such that the solution is continuous at the transition between
the first and second stages of evolution and the constants $A$ and $B$ will be
chosen in order to guarantee that the energy densities of radiation, matter
(baryonic plus dark), and dark energy at the present time are in agreement
with observations.

In section~\ref{sect-numerical-simulations} our numerical results will be
presented in terms of the density parameters for radiation, matter, and dark
energy, the equation-of-state parameter for dark energy, and the deceleration
parameter defined as, respectively,
\begin{eqnarray}
\Omega_{r} &=& \frac{\rho_r}{\rho_c}=\frac{\alpha^2}{6} \rho_{r,0} c^2
e^{-4u}\left( \frac{\dot{a}}{a}\right)^{-2}, \label{densidade-radiation}
\\
\Omega_\textsc{m} &=& \frac{\rho_\textsc{m}}{\rho_c}=\frac{\alpha^2}{6} \left(
B + C e^{\alpha y/2} \right) e^{-3u}\left( \frac{\dot{a}}{a}\right)^{-2},
\label{densidade-matter}
\\
\Omega_y &=& \frac{\rho_y}{\rho_c}= \frac{\alpha^2}{6} \left[
\frac{y_u^2}{2}+Ae^{\alpha y}\left( \frac{\dot{a}}{a}\right)^{-2} \right],
\label{densidade-dark-energy}
\end{eqnarray}
where $\rho_c=3H^2/(8\pi G)=6(\dot{a}/a)^2/(\alpha^2 c^2)$ is the critical
density,
\begin{eqnarray}
w_y=\frac{p_y}{\rho_y c^2}=\frac{y_u^2 \left( \frac{\dot{a}}{a} \right)^2-2 A
e^{\alpha y}}{y_u^2 \left( \frac{\dot{a}}{a} \right)^2+2 A e^{\alpha y}},
\label{equation-state-parameter} \label{wy-u}
\end{eqnarray}
and
\begin{eqnarray}
q=-\frac{a\ddot{a}}{\dot{a}^2}. \label{deceleration-parameter}
\end{eqnarray}

To finish this section, let us point out that a triple unification
of inflation, dark matter, and dark energy was proposed recently
by Liddle, Pahud and
Ure\~{n}a-L\'{o}pez~\cite{liddle-pahud-urena}, using a single
massive scalar field $\phi$ with potential
$V(\phi)=V_0+\frac12m_\phi^2\phi^2$. Note that this potential is
similar to the expanded Salam-Sezgin potential given by
Eq.~(\ref{potencial-expandido}), the main difference being the
fact that the term behaving like dark energy is constant and not
time-dependent as in our case. Within the unification scenario
proposed in Ref.~\cite{liddle-pahud-urena}, a residual inflaton
field $\phi$ survives preheating and oscillates rapidly around the
minimum of the potential, behaving as cold dark matter. However,
similarly to our case, the amplitude of the scalar field
oscillations are too high. They should be drastically reduced in
order to allow for a long enough radiation-dominated epoch
encompassing primordial nucleosynthesis. Exploiting the
uncertainty in the cosmological evolution between the end of the
inflationary period and nucleosynthesis, the authors of
Ref.~\cite{liddle-pahud-urena} considered a brief period of
thermal inflation following preheating, in order to reduce the
amplitude of the scalar field oscillations to the desired level.
In this work, we managed to reduce of the amplitude of the
$x$-field oscillations after inflation within the context of the
Salam-Sezgin cosmological model. As already explained above, this
was accomplished by extending the energy transfer from the
Salam-Sezgin scalar fields to the radiation fluid until well into
the radiation-dominated epoch.

\section{Numerical simulations\label{sect-numerical-simulations}}

Let us now solve numerically the set of equations
(\ref{adotdot1})--(\ref{rho1}) and
(\ref{final-y})--(\ref{alinhaa2}). As initial conditions, at the
beginning of the inflationary period, we choose $x(t_i)=-0.1 \,
m_{\textsc{p}}$ and $y(t_i)=0.1 \, m_{\textsc{p}}$. Note that
enough inflation (65 $e$-folds or more) can be achieved for such
initial amplitudes of the inflaton field $x$, below the Planck
mass, provided that $f_x$ is chosen adequately. In what follows we
choose the phenomenological dissipative parameters to be
$f_x=f_y=275$. In order to achieve agreement with observations,
the constant $A$, introduced in Eq.~(\ref{A}), should be of the
order of $10^{-122} \, m_{\textsc{p}}^4$. We choose, therefore,
$A_1=10^{-12} \, m_{\textsc{p}}^4$, $A_2=1-10^{-110}$, and
$A_3=1$. This choice of $A_1$ also fixes the energy scale of
inflation to be $E_{\mbox{\scriptsize
inf}}=V(x_i,y_i)^{1/4}\approx10^{16}\,\mbox{GeV}$. We assume that,
at the beginning of the inflationary period, the kinetic terms of
the scalar fields, as well as the energy density of radiation, are
of the same order of magnitude of the potential $V$. We choose,
therefore, $\dot{x}(t_i)=\dot{y}(t_i)=1.47\times10^{-6} \,
m_{\textsc{p}}^2$ and $\rho_r(t_i)=4.33\times10^{-12} \,
m_{\textsc{p}}^4$. Finally, we choose $a(t_i)=1$; $\dot{a}(t_i)$
is fixed by the Eq.~(\ref{friedmann1}) to be
$8.52\times10^{-6}m_{\textsc{p}}$.

\begin{figure}[t]
\includegraphics[width=8.6cm]{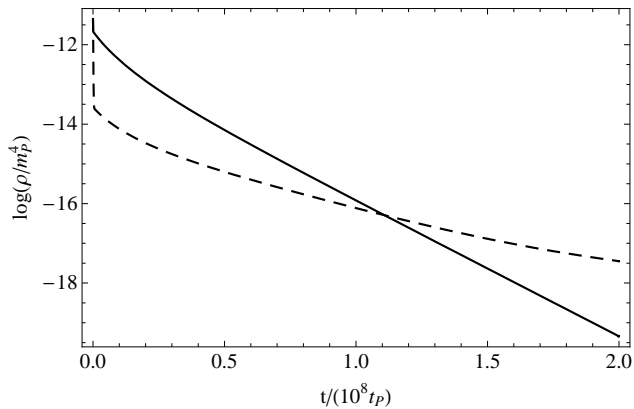}
\caption{Time evolution of the energy densities of radiation
$\rho_r$ (dashed curve) and of the scalar fields $\rho_{xy} =
\dot{x}^2/2 + \dot{y}^2/2 + V(x,y)$ during the first stage of
evolution. The smooth transition from inflation to a
radiation-dominated universe occurs at about $1.1\times10^{8} \,
t_{\textsc{p}}$.} \label{fig-densidades1}
\end{figure}

For such values of the initial conditions and parameters, the
inflationary period lasts for about $1.1\times10^{8} \,
t_{\textsc{p}}$ and the scale factor grows by a factor of about
$10^{28}$. During inflation the condition $\rho_r^{1/4} > H$
holds, as expected in warm inflation. Furthermore, the dissipative
coefficient $\Gamma_x$ is much greater than the Hubble parameter
$H$, implying a strong dissipative warm-inflation regime
\cite{warm-inflation}. At the very beginning of the inflationary
period, the energy density of radiation decreases faster than the
energy density of the scalar fields. However, because of the
energy transfer due to the dissipative terms, the decrease of the
energy density of radiation slows down and, after a while, a
smooth transition from warm inflation to a radiation-dominated
universe takes place (see Fig.~\ref{fig-densidades1}).

\begin{figure}[t]
\includegraphics[width=8.6cm]{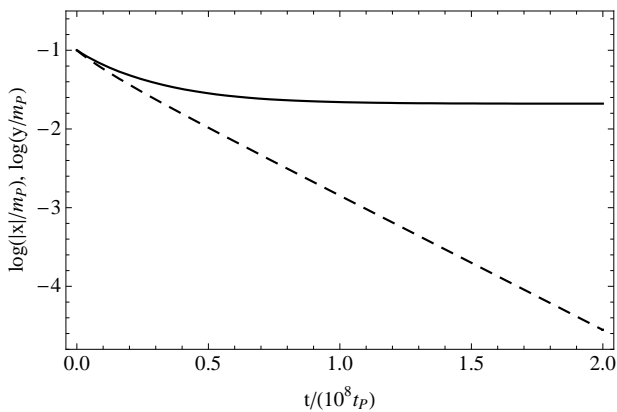}
\caption{Time evolution of the scalar fields $x$ (dashed curve)
and $y$ during the first stage of evolution. After the
inflationary period $y$ decreases very slowly, while the $|x|$
decreases exponentially.} \label{fig-xy1}
\end{figure}

During inflation the $x$ and $y$-fields roll down the Salam-Sezgin
potential. At the end of the inflationary period their values are
$x \approx -9.30 \times10^{-4}\, m_{\textsc{p}}$ and $y\approx
2.16 \times10^{-2}\, m_{\textsc{p}}$. During the subsequent
evolution, till the end of the first stage of evolution, $y$
decreases very slowly, while $|x|$ decreases exponentially (see
Fig.~\ref{fig-xy1}).

As already mentioned above, the residual $x$-field oscillates
rapidly around the minimum of the potential during the second
stage of evolution in which $f_x=f_y=0$, behaving like cold dark
matter. In order to achieve a long radiation-dominated period of
evolution, the initial amplitude of the $x$-field oscillations
should be much smaller than the Planck
mass~\cite{liddle-urena,liddle-pahud-urena}. Our numerical
simulations show that a long enough radiation-dominated epoch
requires, at the time $t_\textsc{t}$ of transition between the
first and second stages of evolution, that $x_\textsc{t}\approx
-10^{-17}\,m_{\textsc{p}}$. Therefore, the energy transfer from
the scalar fields $x$ and $y$ to the radiation fluid should
continue after the end of the inflationary period, allowing $|x|$
to decrease from about $10^{-3}\, m_{\textsc{p}}$, at the end of
inflation, to about $10^{-17}m_{\textsc{p}}$, at the time of
transition between the first and second stages of evolution.

\begin{figure}[t]
\includegraphics[width=8.6cm]{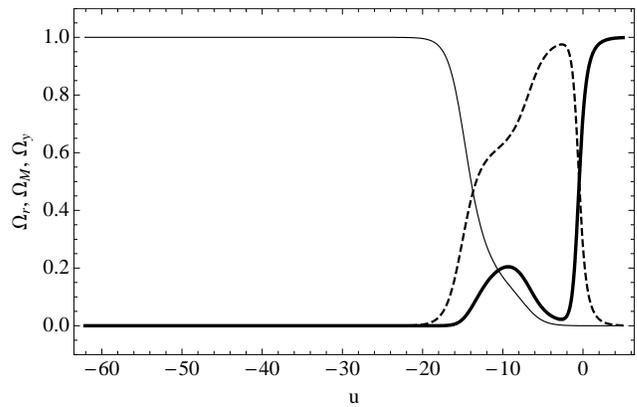}
\caption{Time evolution of the density parameters $\Omega_r$ (thin
curve), $\Omega_\textsc{m}$ (dashed curve), and $\Omega_y$ (thick
curve) during the second stage of evolution. In this case, which
corresponds to a short duration of the first stage of evolution,
$8.00\times10^{8}\, t_{\textsc{p}}$, the transition from a
radiation to a matter-dominated universe occurs earlier in the
history of the universe (but well after nucleosynthesis) and the
density parameter for dark energy becomes non-negligible already
at a redshift $z\approx10^6$. The $x$-field dark matter amounts to
about 10\% of the total matter content of the universe at the
present time $u_0=0$.} \label{fig-dens3}
\end{figure}

The value of $t_\textsc{t}$ should be chosen very carefully. If this value is
too small, the amplitude of the scalar field $x$ does not decrease enough
before it starts to oscillate and, consequently, the transition from a
radiation to a matter-dominated universe takes place too soon. Another
consequence of a shorter first stage of evolution is a rise of the energy
density of the $y$-field (dark energy) at redshift $z\gg1$. In
Fig.~\ref{fig-dens3}, the evolution of radiation, matter (baryonic plus dark),
and dark energy during the second stage of evolution is shown for
$t_\textsc{t}=8.00\times10^{8}\, t_{\textsc{p}}$. In this case, $x_\textsc{t}
= -1.58\times10^{-15}\,m_{\textsc{p}}$ and $\rho_{x,0} = 9.7\times 10^{-2}\,
\rho_{\textsc{m},0}$. On the other hand, if the value of $t_\textsc{t}$ is too
big, the amplitude of the $x$-field decreases too much and its contribution to
the total matter content, at the present time, becomes negligible. For
instance, in the case $t_\textsc{t}=9.50\times10^{8}\, t_{\textsc{p}}$ we have
$x_\textsc{t}=-4.20\times10^{-18}\,m_{\textsc{p}}$ and $\rho_{x,0} =
0.2\times10^{-2}\, \rho_{\textsc{m},0}$.

\begin{figure}[t]
\includegraphics[width=8.6cm]{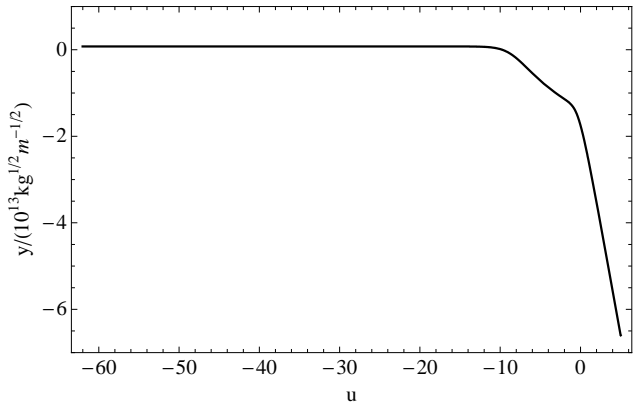}
\caption{Time evolution of the $y$-field dark energy. This field,
which is practically constant during most of the second stage of
evolution, became dominant only in a recent past.} \label{fig-y2}
\end{figure}

\begin{figure}[t]
\includegraphics[width=8.6cm]{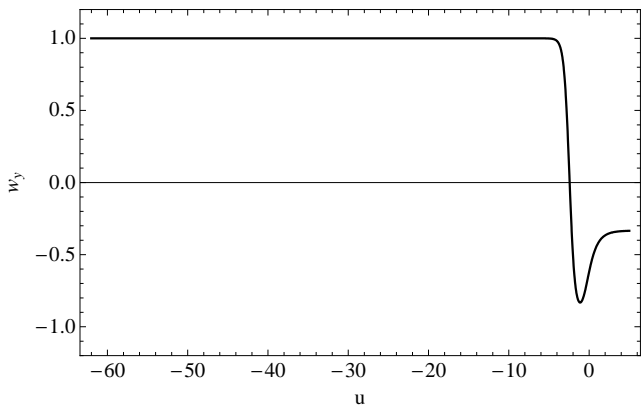}
\caption{Time evolution of the equation-of-state parameter for
dark energy, $w_y$, during the second stage of evolution. At the
present time, $u_0=0$, the value of this parameter is about
$-0.62$.  In the future, it will approach the value $-1/3$.}
\label{fig-wy2}
\end{figure}

\begin{figure}[t]
\includegraphics[width=8.6cm]{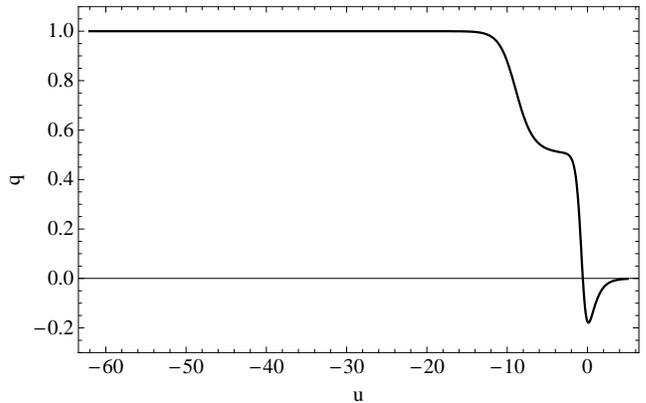}
\caption{Time evolution of the deceleration parameter, $q$, during
the second stage of evolution. This parameter has its minimum
(negative) value around the present time and approaches zero in
the future.} \label{fig-q2}
\end{figure}

\begin{figure}[t]
\includegraphics[width=8.6cm]{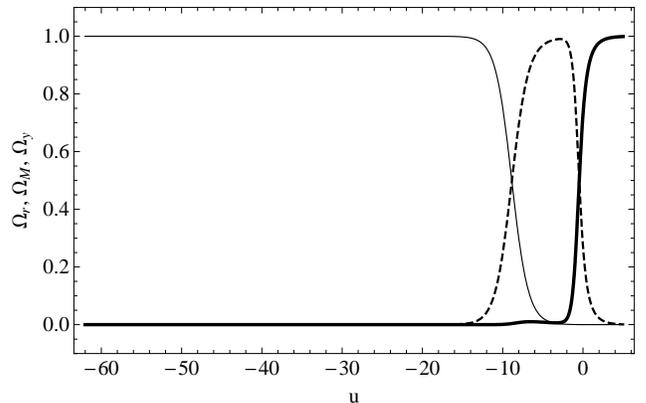}
\caption{Time evolution of the density parameters $\Omega_r$ (thin
curve), $\Omega_\textsc{m}$ (dashed curve), and $\Omega_y$ (thick
curve) during the second stage of evolution. In this case, which
corresponds to a duration of the first stage of evolution of
$8.90\times10^{8}\, t_{\textsc{p}}$, the density parameter for
dark energy becomes non-negligible only recently. At the present
time, $u_0=0$, the density parameters for radiation,
$\Omega_{r,0}=4.8\times10^{-5}$, matter,
$\Omega_{\textsc{m},0}=0.27$, and dark energy,
$\Omega_{y,0}=0.73$, are in good agreement with observational
data. The $x$-field dark matter contributes only about 6\% to the
total matter content of the universe.} \label{fig-dens2}
\end{figure}

Let us now analyze in more detail the solution obtained for
$t_\textsc{t}=8.90\times10^{8}\, t_{\textsc{p}}$. In this case, at
the time of transition between the first and second stages of
evolution, we have $a_\textsc{t}=9.85 \times 10^{28}$,
$x_\textsc{t}=-4.45 \times 10^{-17}\, m_{\textsc{p}}$,
$y_\textsc{t}=2.09\times 10^{-2} \, m_{\textsc{p}}$,
$\rho_{x,\textsc{t}}= 1.16\times 10^{-43}\, m_{\textsc{p}}^4 =
5.96 \times 10^{53}\,\mbox{kg/m}^3$, and
$\rho_{r,\textsc{t}}=4.89\times10^{-20}\, m_{\textsc{p}}^4 = 2.52
\times 10^{77}\,\mbox{kg/m}^3$. Taking into account that
$\rho_{r,0} = 8.92\times 10^{-128}\, m_{\textsc{p}}^4= 4.6 \times
10^{-31}\,\mbox{kg/m}^3$, we obtain $u_\textsc{t} = (1/4)
\ln(\rho_{r,0}/\rho_{r,\textsc{t}}) = -62.02$, and, from
Eq.~(\ref{fix-C}), $C=1.68\times 10^{-124}\, m_{\textsc{p}}^4=7.80
\times 10^{-11}\,\mbox{kg\,m}^{-1}\mbox{s}^{-2}$. The constants
$A$ and $B$ are chosen in order to satisfy the observational
constraints (\ref{constraint-A}) and (\ref{constraint-B}), namely,
$A=3.16\times 10^{-122}\, m_{\textsc{p}}^4 = 1.47 \times
10^{-8}\,\mbox{kg\,m}^{-1}\mbox{s}^{-2}$ and $B=4.73\times
10^{-124}\, m_{\textsc{p}}^4 = 2.19 \times 10^{-10}
\,\mbox{kg.m}^{-1}\mbox{s}^{-2}$. In this case, we obtain that the
$x$-field dark matter contributes about $6\%$ to the total matter
content of the universe at the present time,
$\rho_{x,0}=6.15\times10^{-2}\, \rho_{\textsc{m},0}$. The time
evolution of the quantities $y$, $w_y$, $q$, $\Omega_r$,
$\Omega_\textsc{m}$, and $\Omega_y$, defined in
Eqs.~(\ref{densidade-radiation})--(\ref{deceleration-parameter}),
is shown in Figs.~\ref{fig-y2}--\ref{fig-dens2} for the case under
consideration.

\begin{figure}[t]
\includegraphics[width=8.6cm]{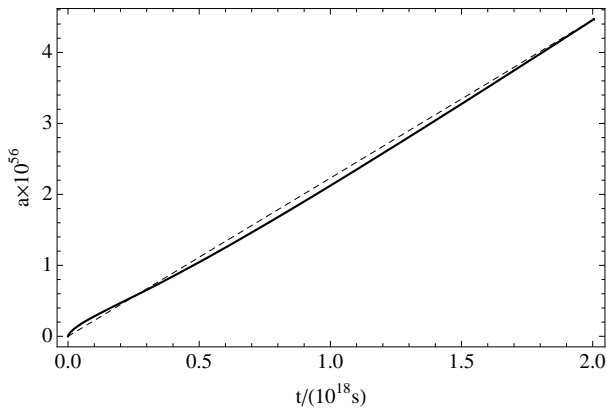}
\caption{Time evolution of the scale factor $a$ (thick curve). To
guide the eye, a linear growth is also represented (dashed
straight line). The transition to an accelerated expansion takes
place at about $(2-3)\times10^{17}\,\mbox{s}$. At the present
time, $t\approx 4\times 10^{17}\,\mbox{s}$, the expansion of the
universe is slightly accelerated. In the future, the scale factor
growth approaches a linear behavior, a consequence of the fact
that the equation-of-state parameter for dark energy approaches
$-1/3$.} \label{fig-factor-de-escala}
\end{figure}

To finish this section, let us comment on the evolution of the
scale factor today and in the future. As can be seen in
Fig.~\ref{fig-wy2}, the equation-of-state parameter for dark
matter, $w_y$, is today about $-0.62$, implying that $a(t)\propto
t^{1.75}$, if the matter and radiation contributions are
neglected. However, there is today a non-negligible contribution
from matter ($\Omega_{\textsc{m},0}=0.27$), which if taken alone
would imply $a(t)\propto t^{2/3}$. Taking into account both
contributions, from dark energy and from matter, we obtain that
today the expansion of the universe is slightly accelerated.
Naively it would seem that this acceleration should increase in
the future, since the density parameter for matter will approach
zero and dark energy will dominate the dynamics of evolution of
the universe. However, at the same time as
$\Omega_{\textsc{m}}\rightarrow 0$, the equation-of-state
parameter for dark energy, $w_y$, approaches $-1/3$ (see
Fig.~\ref{fig-wy2}). As a result, the deceleration parameter, $q$,
approaches zero (see Fig.~\ref{fig-q2}) and the scale factor grows
linearly in the future. In Fig.~\ref{fig-factor-de-escala} we show
the evolution of the scale factor into the future, obtained by
direct numerical integration of Eqs.~(\ref{ydotdot}) and
(\ref{eq-adota}), using the same initial conditions and values of
the parameters $A$, $B$, and $C$ as in
Figs.~\ref{fig-y2}--\ref{fig-dens2}.

\section{Gravitational waves\label{sect-gravitational-waves}}

In this section, we calculate the full gravitational-wave spectrum of the
Salam-Sezgin cosmological model using the formalism of the continuous
Bogoliubov coefficients. This formalism was first applied to particle
production in an expanding universe by Parker~\cite{parker} and later extended
to the case of gravitons by Henriques~\cite{henriques1} and Moorhouse,
Henriques, and Mendes~\cite{moorhouse-henriques-mendes}. It avoids, in a
natural way, overproduction of gravitons of high
frequencies~\cite{mendes-henriques-moorhouse}. In recent years, the formalism
of continuously evolving Bogoliubov coefficients was applied to the
calculation of gravitational-wave spectra in several cosmological models,
revealing interesting features in the MHz/GHz range of frequencies due to the
transition between the inflationary and the radiation-dominated
eras~\cite{henriques,sa-henriques-1,sa-henriques-2}.

Let us define the gravitational-wave spectral energy density
parameter as
\begin{eqnarray}
\Omega_\textsc{gw} \equiv  \frac{1}{\rho_c}
\frac{d\rho_\textsc{gw}}{d \ln\omega} \nonumber = \frac{8\hbar
G}{3\pi c^5 H^2} \omega^4 \beta^2, \label{sedp}
\end{eqnarray}
where $\rho_\textsc{gw}c^2$ and $\omega$ are the energy density
and angular frequency of the gravitational waves, respectively,
$\rho_c c^2$ is the critical energy density of the universe, $H$
is the Hubble parameter, and $\beta$ is a Bogoliubov coefficient,
such that $\beta^2$ gives the number of gravitons. All quantities
in the above expressions are evaluated at the present time,
$u_0=0$. The squared Bogoliubov coefficient $\beta^2$ is given by
$\beta^2=(X-Y)^2/4$, where $X$ and $Y$ are continuous functions of
time, determined during the first stage of evolution by the set of
differential equations
\begin{eqnarray}
 \dot{X} &=& - i \omega_0 \left( \frac{a_0}{a} \right)Y, \label{XX1}
\\
 \dot{Y} &=& -\frac{i}{\omega_0} \frac{a}{a_0} \left[ \omega_0^2
\left( \frac{a_0}{a} \right)^2 - \frac{\ddot{a}}{a}- \left(
\frac{\dot{a}}{a} \right)^2\right]X, \label{YY1}
\end{eqnarray}
where $\ddot{a}/a$ and $(\dot{a}/a)^2$ are given by
Eqs.~(\ref{adotdot1}) and (\ref{friedmann1}), respectively, and
during the second stage of evolution by the set of differential
equations
\begin{eqnarray}
\hspace{-2mm} X_u &=& - i \omega_0 \left( \frac{a_0}{a} \right)
\frac{Y}{\dot{a}/a}, \label{XX2}
\\
\hspace{-2mm} Y_u &=& -\frac{i}{\omega_0} \frac{a}{a_0} \left[
\omega_0^2 \left( \frac{a_0}{a} \right)^2 - \frac{\ddot{a}}{a}-
\left( \frac{\dot{a}}{a} \right)^2\right] \frac{X}{\dot{a}/a},
\label{YY2}
\end{eqnarray}
where $\ddot{a}/a$ and $(\dot{a}/a)^2$ are given by
Eqs.~(\ref{alinhalinhaa}) and (\ref{alinhaa2}), respectively. The
scale factor at the present time is given by
$a_0=a_\textsc{t}e^{-u_\textsc{t}}$, where the subscript
$\textsc{t}$ stands for the time of transition between the first
and second stages of evolution.

The above differential equations are integrated numerically, from
the beginning of the inflationary period till the present time. We
assume that prior to the inflationary period the universe was
radiation-dominated and, therefore, no gravitons were created,
i.e., $\alpha^2(t)=1$ and $\beta^2(t)=0$ for $t \leq t_i$, where
$\alpha$ is a Bogoliubov coefficient related to $X$ and $Y$
through $\alpha^2=(X+Y)^2/4$. The appropriate initial conditions
for the functions $X$ and $Y$ are then $X(t_i)=1$ and $Y(t_i)=1$.
The gravitational-wave angular frequency at present time,
$\omega_0$, is taken to vary from about $10^{-17}\,\mbox{rad/s}$
(corresponding to a wavelength equal, today, to the Hubble
distance) to about $10^{8}\,\mbox{rad/s}$ (corresponding to a
wavelength equal to the Hubble distance at the end of the
inflationary period).

\begin{figure}[t]
\includegraphics[width=8.6cm]{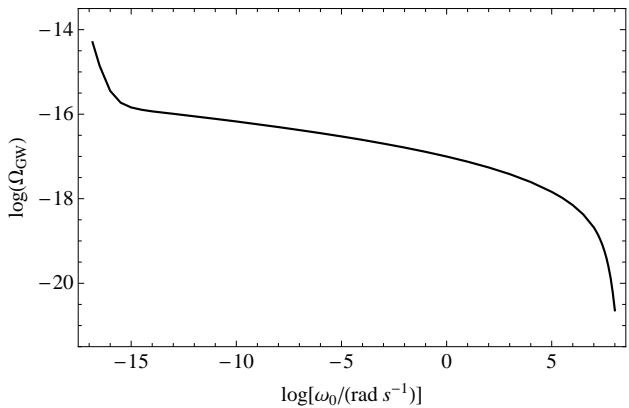}
\caption{Gravitational wave spectrum of the Salam-Sezgin
cosmological model.} \label{fig-espectro}
\end{figure}

The full gravitational-wave spectrum for the Salam-Sezgin
cosmological model is shown in Fig.~\ref{fig-espectro}. It was
calculated for the particular case of the model with
$t_\textsc{t}=8.90\times10^{8}\, t_{\textsc{p}}$, discussed in
detail at the end of Sect.~\ref{sect-numerical-simulations}.

For angular frequencies corresponding to the present size of the
Hubble horizon, an upper limit on the gravitational-wave spectral
energy density parameter can be derived from measurements of the
cosmic microwave background radiation, namely,
$\Omega_{\textsc{gw}} < 1.4 \times 10^{-10}$ for
$\omega_{\mbox{\scriptsize hor}}=1.4\times10^{-17}\,\mbox{rad/s}$
\cite{allen}. The gravitational-wave spectrum of the Salam-Sezgin
cosmological model satisfies this constraint easily.

\begin{figure}[t]
\includegraphics[width=8.6cm]{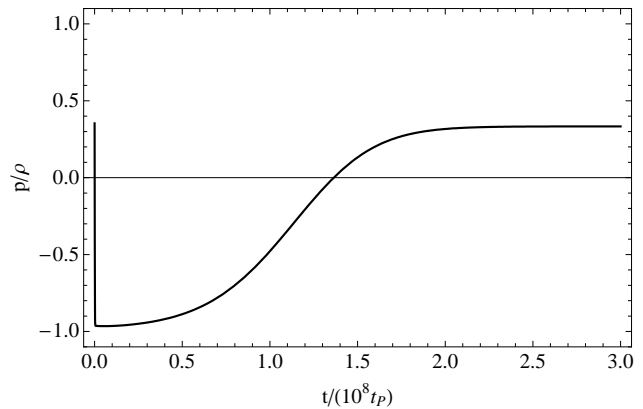}
\caption{Time evolution of the equation-of-state parameter,
$w=p/\rho$, during the first stage of evolution.}
\label{fig-equacao-estado-1}
\end{figure}

In the intermediate range of frequencies, from about
$10^{-15}\,\mbox{rad/s}$ to about $10^{6}\,\mbox{rad/s}$, the
gravitational-wave spectrum differs substantially from the ones
obtained, with the same formalism, in
Refs.~\cite{henriques,sa-henriques-1,sa-henriques-2}, in that it
has a non-constant slope. This is due to the fact that, in
previous investigations, numerical evaluation of the Bogoliubov
coefficients started at the end of inflation and, moreover, it was
assumed, for the purpose of determining the initial conditions for
the functions $X$ and $Y$, that the inflationary period was either
exponential or of the power-law type. This assumption was needed
since exact analytical solutions of Eqs.~(\ref{XX1}) and
(\ref{YY1}) are known only for these types of inflation. As a
result, the gravitational-wave spectrum had, in the intermediate
range of frequencies, zero slope~\cite{henriques,sa-henriques-1}
or constant negative slope~\cite{sa-henriques-2}, respectively. In
this article, we have improved our calculation by numerically
evaluating the Bogoliubov coefficients from the very beginning of
the inflationary period. This approach allow us to consider any
type of inflation, not just exponential or power-law. This is
rather convenient for the cosmological model under consideration,
since, during inflation, the equation-of-state parameter
$w=p/\rho$ changes smoothly from $-1$ to $-1/3$ (see
Fig.~\ref{fig-equacao-estado-1}), i.e., inflation is neither
exponential nor power-law.

The shape of the spectrum in the intermediate range of frequencies
can be understood as follows. It is known that, for power-law
inflationary models, the spectrum has constant negative slope in
this range of frequencies~\cite{sahni} (see also
Ref.~\cite{sa-henriques-2} for the derivation of this result using
the formalism of continuous Bogoliubov coefficients). In the limit
where power-law inflation approaches exponential inflation
($s\rightarrow +\infty$, where $a(t)\propto t^s$), the slope of
the spectrum approaches zero. Now, let us regard the evolution of
our cosmological model, during inflation, as a succession of short
periods of time, with duration $\Delta t$, such that in each of
them inflation can be considered to be of the power-law type with
$s=\frac23(w+1)^{-1}\thickapprox\mbox{constant}$. Gravitational
waves that cross the Hubble horizon during the first of these
periods, have today low frequencies. Since during this period the
equation-of-state parameter is $w=p/\rho\thickapprox-1$ (see
Fig.~\ref{fig-equacao-estado-1}), corresponding to $s\gg1$, the
spectrum, for such low frequencies, is almost flat. During the
second period of time, $w$ is slightly higher and the spectrum has
a small negative slope in the corresponding frequency range.
Continuing with this process, we obtain a spectrum that is a
succession of line segments with increasing (in modulus) slope as
the frequency of the gravitational waves increases. Taking the
limit $\Delta t \rightarrow 0$, we obtain, in the intermediate
region of the spectrum, the shape shown in
Fig.~\ref{fig-espectro}.

The gravitational-wave spectrum of the Salam-Sezgin cosmological
model shows no structure in the MHz/GHz frequency range. Such a
structure is expected in cosmological models in which standard
(cold) inflation is followed by a preheating and/or reheating
stage~\cite{henriques,sa-henriques-1,sa-henriques-2}. In such
models, after the inflationary period the inflaton field
oscillates about the minimum of the its potential, leading to
particle production and an increase of the energy density of
radiation. The structure of the gravitational-wave spectrum in the
MHZ/GHz range of frequencies depends on the form of the
inflationary potential near its minimum. More specifically, if the
potential near its minimum is proportional to $\phi^n$, then
$\Omega_\textsc{gw}$ increases with the increase of
$n$~\cite{sa-henriques-2}. Contrarily to these standard (cold)
inflationary cosmological models, we have assumed that, within the
Salam-Sezgin cosmological model, inflation is of the warm type.
That is, the energy density of the $x$ and $y$-fields is
continuously transferred to the radiation fluid and, consequently,
the energy density of radiation is not diluted during inflation.
Because the energy density of radiation decreases slower than the
energy density of the $x$ and $y$-fields (see
Fig.~\ref{fig-densidades1}), the transition from inflation to the
radiation-dominated era occurs smoothly, well before the $x$-field
starts to oscillate about the minimum of its potential. Therefore,
the coherent oscillations of the $x$-field leave no imprint in
gravitational-wave spectrum in the MHz/GHz range of frequencies.

\section{Conclusions\label{sect-conclusions}}

In this work we have extended the analysis by Anchordoqui,
Goldberg, Nawata and Nu\~nez based on the Salam-Sezgin
supergravity theory, by including the inflationary period in the
evolution. The Salam-Sezgin potential contains two scalar fields,
$x$ and $y$. Assuming sufficient dissipative coupling, it follows
that a period of warm inflation takes place, driven by the
$x$-field, during which energy is being transferred continuously
from both scalars to a radiation fluid. Well after the end of the
inflationary period, with the universe already dominated by the
energy density of the radiation fluid, the dissipative
coefficients are switched off and the $x$-field begins a period of
quick oscillations around the minimum of the potential, behaving
like cold dark matter with a varying mass. The $y$-field, which
behaves like dark energy, becomes then practically constant until
very near the present epoch, when it begins to dominate the energy
content of the universe. We have shown, therefore, that within the
Salam-Sezgin cosmological model is possible to obtain a unified
description of inflation, dark matter, and dark energy. For an
appropriate choice of the parameters of the model, we were able to
account for the observed values of radiation, matter, and
dark-energy density parameters. However, only approximately 6\% of
the total matter content of the universe was obtained from the
$x$-field, while, according to observational data, one would
expect dark matter to amount to about 80\% of the total matter
content. Even if not entirely successful in this respect, it is
not unreasonable to expect that more sophisticated models, based
on supergravity and superstring theories, could be found that may
generate enough dark matter and dark energy in a fully unified
description.

We have also calculated the full gravitational-wave spectrum of
the Salam-Sezgin cosmological model. Using the formalism of the
continuous Bogoliubov coefficients, we have shown that this
spectrum has a non-constant negative slope in the frequency range
$(10^{-15}-10^{6})\,\mbox{rad/s}$, and that, contrarily to
cosmological models in which standard (cold) inflation is followed
by a preheating and/or reheating stage, the gravitational-wave
spectrum shows no structure in the MHz/GHz range of frequencies.

\begin{acknowledgments}
This work was supported in part by the Funda\c{c}\~ao para a
Ci\^encia e a Tecnologia, Portugal.
\end{acknowledgments}

\end{document}